\documentstyle[12pt,psfig,epsf]{article}
\newcommand{\be}{\begin{equation}}
\newcommand{\ee}{\end{equation}}

\newcommand{\beqa}{\begin{eqnarray}}
\newcommand{\eeqa}{\end{eqnarray}}

\newcommand{\eqref}[1]{(\ref{#1})}

\def\boxit#1{\vbox{\hrule\hbox{\vrule\kern8pt
\vbox{\hbox{\kern8pt}\hbox{\vbox{#1}}\hbox{\kern8pt}}
\kern8pt\vrule}\hrule}}
\def\mathboxit#1{\vbox{\hrule\hbox{\vrule\kern8pt\vbox{\kern8pt
\hbox{$\displaystyle #1$}\kern8pt}\kern8pt\vrule}\hrule}}

\def\IB{\relax\hbox{$\inbar\kern-.3em{\rm B}$}}
\def\IC{\relax\hbox{$\inbar\kern-.3em{\rm C}$}}
\def\ID{\relax\hbox{$\inbar\kern-.3em{\rm D}$}}
\def\IE{\relax\hbox{$\inbar\kern-.3em{\rm E}$}}
\def\IF{\relax\hbox{$\inbar\kern-.3em{\rm F}$}}
\def\IG{\relax\hbox{$\inbar\kern-.3em{\rm G}$}}
\def\IGa{\relax\hbox{${\rm I}\kern-.18em\Gamma$}}
\def\IH{\relax{\rm I\kern-.18em H}}
\def\IK{\relax{\rm I\kern-.18em K}}
\def\IL{\relax{\rm I\kern-.18em L}}
\def\IP{\relax{\rm I\kern-.18em P}}
\def\IR{\relax{\rm I\kern-.18em R}}
\def\IZ{\relax\ifmmode\mathchoice
{\hbox{\cmss Z\kern-.4em Z}}{\hbox{\cmss Z\kern-.4em Z}}
{\lower.9pt\hbox{\cmsss Z\kern-.4em Z}} {\lower1.2pt\hbox{\cmsss
Z\kern-.4em Z}}\else{\cmss Z\kern-.4em Z}\fi}

\def\II{\relax{\rm I\kern-.18em I}}

\begin{document}

\hfill  NRCPS-HE-02-23

\vspace{5cm}
\begin{center}
{\LARGE New strings  \\ with \\world-sheet
supersymmetry

}

\vspace{2cm}

{\sl A.Nichols\footnote{nichols@inp.demokritos.gr},
R.Manvelyan\footnote{manvel@moon.yerphi.am},
G.K.Savvidy\footnote{savvidy@inp.demokritos.gr\\
}\\
National Research Center Demokritos,\\
Ag. Paraskevi, GR-15310 Athens, Hellenic Republic\\

}
\end{center}
\vspace{60pt}

\centerline{{\bf Abstract}}

\vspace{12pt}

\noindent We suggest a new model of string
theory with world-sheet supersymmetry.
It possesses an additional global fermionic symmetry which
is similar in many ways to BRST symmetry.
The spectrum consists of massless states of Rarita-Schwinger fields
describing infinite tower of half-integer spins.


\newpage

\pagestyle{plain}

\section{Introduction}

In the recent articles \cite{gon1,gon3} there was suggested
a string theory which is described by the following action:
\be\label{funaction}
S =m \cdot L= {m\over \pi} \int d^{2}\zeta
\sqrt{h}\sqrt{K^{ia}_{a}K^{ib}_{b}},
\ee
here $m$ has dimension of mass, $h_{ab}$ is the induced metric and
$K^{i}_{ab}$ is the second fundamental form (extrinsic curvature)
\footnote{This action is essentially different
in its geometrical meaning from the action considered in
previous studies \cite{polykov} where it is proportional
to the spherical angle and has dimensionless coupling constant.}.
Instead of being proportional to the area of the surfaces, as it
is the case in the standard string theory,
the action (\ref{funaction}) is proportional
to the length of the surface $ L $ \cite{gon1}.
Due to the last property the model
has two essentially new properties, first of all, when the surface degenerates
into a single world line, the action (\ref{funaction}) becomes
proportional to the length of the world line
\be\label{limit}
S= m ~A_{xy} ~~ \rightarrow ~~  m~\int^{Y}_{X} ds
\ee
and the functional integral over surfaces naturally transforms
into the Feynman path integral for a point-like relativistic particle,
thus naturally extending it to relativistic strings and, secondly,
the action is equal to the perimeter of the Wilson loop $S=m(R+T)$,
where R is space distance between quarks,
therefore at the {\it classical level string tension is equal to zero}.
Quantization of this system and its massless spectrum
have been derived in \cite{gon3}. In this string theory
{\it all particles, with arbitrary large integer spin, are massless}. This
pure massless spectrum is consistent with the tensionless character
of the model and it was conjectured in \cite{gon3}
that it may describe unbroken phase of standard string theory  when
$ \alpha^{'} \rightarrow \infty$ and all masses tend to zero
$M^{2}_{n} = {1\over \alpha^{'}}(n-1) \rightarrow 0$ \cite{gross} .

Our aim now is to introduce fermions   and to suggest
supersymmetric extension of this model using world-sheet
superfields \cite{ramond,neveu,zumino,green,polchinski}.
The action(\ref{funaction}) can
be written in the equivalent from \cite{gon1,gon3}
\begin{equation}\label{conaction}
S= {m\over\pi}\int d^{2}\zeta \sqrt{h}\sqrt{ \left(\Delta(h)
X_{\mu}\right)^{2}},
\end{equation}
here ~$h_{ab}=\partial_{a}X_{\mu}\partial_{b}X_{\mu}$ ~is the induced
metric,~$\Delta(h)= 1/\sqrt{h}~\partial_{a}\sqrt{h}h^{ab}
\partial_{b} $ ~is Laplace operator, \footnote{The equivalence follows
from the relation:
$K^{ia}_{a}K^{ib}_{b}=\left(\Delta(h) X_{\mu}\right)^{2},
\quad i,j=1,2,...,D-2$}
$a,b=1,2; \qquad \mu=0,1,2,...,D-1$.
We shall fix the conformal gauge $h_{ab}=\rho\eta_{ab}$ using the
reparametrization invariance of the action (\ref{conaction})
and represent it in two equivalent forms  \cite{gon3}
\be\label{gaga}
S={m\over\pi} \int d^{2}\zeta \sqrt{\left(\partial^{2}
X \right)^{2}} ~~\Leftrightarrow~~\acute{S} =
{1\over\pi}\int d^{2}\zeta \{~\Pi^{\mu}~\partial^{2} X^{\mu}
- 2 \Omega ~ ( \Pi^{2} -m^2)~\},
\ee
where the independent field $\Pi^{\mu}$ and the Lagrange multiplier  $\Omega$
have been introduced. The system of equations which follows from
$\acute{S}$
\beqa\label{orig}
\partial^2 \Pi^{\mu} =0,~~~~~~~~
\partial^2 X^{\mu} - 2\Omega \Pi^{\mu} =0,~~~~~~~~
\Pi^{\mu}\Pi_{\mu}~ = m^{2}
\eeqa
is equivalent to the original equation for $X^{\mu}$ and
the $\Pi^{\mu}$ field takes the form
$$\Pi^{\mu} = m \frac{\partial^{2}X^{\mu}}
{\sqrt{\left(\partial^{2}X \right)^2}}.
$$
Both forms of the action (\ref{gaga}) can be extended to the supersymmetric
case as follows.

\section{N=1 World-sheet Supersymmetry}
For the basic fields~$(X,\Pi,\Omega)$ in (\ref{gaga})
we shall introduce the corresponding
superfields \cite{zumino,green,polchinski}
\beqa
\hat{X}^{\mu} &=& X^{\mu} + \bar{\vartheta} \Psi^{\mu}+ {1\over 2}
\bar{\vartheta} \vartheta F^{\mu}\nonumber\\
\hat{\Pi}^{\mu} &=& \Pi^{\mu} + \bar{\vartheta} \eta^{\mu}+ {1\over 2}
\bar{\vartheta} \vartheta \Phi^{\mu}\nonumber\\
\hat{\Omega}  &=&  \omega  + \bar{\vartheta} \xi + {1\over 2}
\bar{\vartheta} \vartheta \Omega ,
\eeqa
where $\vartheta$ is an anti-commuting variable
and shall define the supersymmetric action
simply exchanging basic fields ~$(X,\Pi,\Omega)$
in (\ref{gaga}) by corresponding superfields
as follows
\be\label{firstordsusy}
S = {-i \over 2\pi} \int d^{2}\zeta d^{2} \theta \{ ~\hat{\Pi}^{\mu}
\bar{D} D \hat{X}^{\mu} - 2\hat{\Omega} ( \hat{\Pi}^{2} -m^2)~\},
\ee
where
\be
D_A = { \partial \over \partial \bar{\vartheta}^A} -
i (\rho^{a} \vartheta)_A \partial_{a},~~~~
\Psi^{\mu}_{A}(\zeta) \equiv \left( \begin{array}{c}
     \Psi^{\mu}_{-}(\zeta)\\
     \Psi^{\mu}_{+}(\zeta)
     \end{array} \right),~\eta^{\mu}_{A}(\zeta)
     \equiv \left( \begin{array}{c}
     \eta^{\mu}_{-}(\zeta)\\
     \eta^{\mu}_{+}(\zeta)
     \end{array} \right),~\xi_{A}(\zeta)
     \equiv \left( \begin{array}{c}
     \xi_{-}(\zeta)\\
     \xi_{+}(\zeta)
     \end{array} \right),
\ee
$\mu$ is a space-time vector index, $A=1,2$ is a
two-dimensional spinor index.
 $\bar{\Psi}^{\mu} = \Psi^{+\mu} \rho^{0}= \Psi^{T\mu} \rho^{0}$ and
$\rho^{\alpha}$ are two-dimensional Dirac matrices
\be
\{ \rho^{a},\rho^{b} \} =-2 \eta^{a b}.
\ee
In Majorana basis the $\rho's$ are given by
\be
\rho^{0} = \left( \begin{array}{cc}
     0&-i\\
     i&0
     \end{array} \right),~~~~\rho^{1} = \left( \begin{array}{cc}
     0&i\\
     i&0
     \end{array} \right)
\ee
and $i\rho^{\alpha}\partial_{\alpha}$ is a real operator.
The two-dimensional chiral fields are defined as
$
\rho^{3}\Psi^{\mu}_{\pm} = \mp \Psi^{\mu}_{\pm},
$
where $\rho^{3}=\rho^{0}\rho^{1}$. We should compute different
expressions involved in the action,
\beqa
\bar{D}^{A} D_A \hat{X}^{\mu} = 2 F^{\mu} +
2i \bar{\vartheta} \rho^{a} \partial_{a}\Psi^{\mu} -
\bar{\vartheta} \vartheta \partial^2 X^{\mu}, \nonumber
\eeqa
thus
\beqa
\hat{\Pi}^{\mu} \bar{D} D \hat{X}^{\mu} = (\Pi^{\mu} + \bar{\vartheta} \eta^{\mu}+ {1\over 2}
\bar{\vartheta} \vartheta \Phi^{\mu})(2 F^{\mu} +
2i \bar{\vartheta} \rho^{a} \partial_{a}\Psi^{\mu} -
\bar{\vartheta} \vartheta \partial^2 X^{\mu}) \nonumber
\eeqa
and the quadratic part in $\vartheta$ is equal to
\beqa
-\bar{\vartheta} \vartheta ~ \Pi^{\mu}\partial^2 X^{\mu} +
\bar{\vartheta} \vartheta ~F^{\mu}\Phi^{\mu} +
2i \bar{\vartheta} \eta^{\mu}~\bar{\vartheta} \rho^{a}
\partial_{a}\Psi^{\mu}.
\nonumber
\eeqa
The integral over Grassmann variables is defined as
$\int d^{2} \theta \bar{\vartheta} \vartheta = -2i$,
therefore
\beqa
{-i \over 2} \int d^{2} \theta \{ ~-\bar{\vartheta}
\vartheta ~ \Pi^{\mu}\partial^2 X^{\mu} +
\bar{\vartheta} \vartheta ~F^{\mu}\Phi^{\mu} +
2i \bar{\vartheta} \eta^{\mu}~\bar{\vartheta} \rho^{a}
\partial_{a}\Psi^{\mu}~\} \nonumber
\eeqa
\beqa
=\Pi^{\mu}\partial^2 X^{\mu} +
i\bar{\eta^{\mu}}\rho^{a} \partial_{a}\Psi^{\mu}
-F^{\mu}\Phi^{\mu},\nonumber
\eeqa
where we have used the relation
$$
2i \bar{\vartheta} \eta^{\mu}~\bar{\vartheta} \rho^{a} \partial_{a}\Psi^{\mu}=
-i \bar{\vartheta} \vartheta \bar{\eta^{\mu}}\rho^{a} \partial_{a}\Psi^{\mu}.
\nonumber
$$
Now we have to compute also the second term in the Lagrangian
\beqa
\hat{\Omega}  ( \hat{\Pi}^{2} -m^2) = \omega (\Pi^2 -m^2)
+2 \omega \Pi^{\mu}\bar{\vartheta} \eta^{\mu} + \omega(\bar{\vartheta} \eta^{\mu}
\bar{\vartheta} \eta^{\mu} +\bar{\vartheta} \vartheta ~\Pi^{\mu}\Phi^{\mu})
\nonumber\\
+ \bar{\vartheta}\xi (\Pi^2 -m^2)  + 2 \bar{\vartheta}\xi
\Pi^{\mu}\bar{\vartheta} \eta^{\mu} + {1\over 2}
\bar{\vartheta} \vartheta \Omega (\Pi^2 -m^2) \nonumber
\eeqa
and to integrate it over Grassmann variables
\beqa
{-i \over 2} \int d^{2} \theta \{-2\hat{\Omega}  ( \hat{\Pi}^{2} -m^2)\}
=\Omega (\Pi^2 -m^2) +  \omega (2\Pi^{\mu}\Phi^{\mu} +
\bar{\eta^{\mu}}\eta^{\mu}) + 2 \Pi^{\mu}~\bar{\eta^{\mu}}\xi. \nonumber
\eeqa
The full action is now equal  to the following expression
\beqa\label{fullaction}
S =
{1\over \pi} \int d^{2}\zeta ~\{ ~\Pi^{\mu}\partial^2 X^{\mu}
+ i\bar{\eta^{\mu}}\rho^{a} \partial_{a}\Psi^{\mu}
-F^{\mu}\Phi^{\mu}~~\nonumber\\
~~-\Omega (\Pi^2 -m^2)
-  \omega (2\Pi^{\mu}\Phi^{\mu} +
\bar{\eta^{\mu}}\eta^{\mu}) - 2 \Pi^{\mu}~\bar{\eta^{\mu}}\xi~\}.
\eeqa
The equations of motion are:
\beqa\label{fullequationa}
(I)~~~~~~~~~~~~~~~~~~~\Phi^{\mu} =0\nonumber\\
\partial^2 \Pi^{\mu} =0\nonumber\\
i \rho^{a} \partial_{a}\eta^{\mu}=0\nonumber\\
2 \omega \Pi^{\mu} + F^{\mu} =0\nonumber\\
\partial^2 X^{\mu} - 2\Omega \Pi^{\mu} -2\omega \Phi^{\mu}
-2\bar{\eta^{\mu}}\xi=0\nonumber\\
i\rho^{a} \partial_{a}\Psi^{\mu} -2\Pi^{\mu}\xi -2 \omega\eta^{\mu}=0,
\eeqa
and the variation over Lagrange multipliers gives constraints
\beqa\label{fullconstrants}
(II)~~~~~~~~~~~~~~~~~~\Pi^2 -m^2 =0\nonumber\\
2\Pi^{\mu}\Phi^{\mu} +
\bar{\eta^{\mu}}\eta^{\mu}=0\nonumber\\
2 \Pi^{\mu}~\eta^{\mu} =0.
\eeqa
The SUSY transformation is:
\beqa
\begin{array}{lll}\label{susy}
\delta X^{\mu} = \bar{\epsilon}\Psi^{\mu},\\
\delta \Psi^{\mu} = -i \rho^{a} \partial_{a}
X^{\mu} ~\epsilon  ~+~F^{\mu}~\epsilon ,\\
\delta F^{\mu} = -i \bar{\epsilon} \rho^{a} \partial_{a}\Psi^{\mu},
\end{array}
\begin{array}{lll}
\delta \Pi^{\mu} = \bar{\epsilon}\eta^{\mu},\\
\delta \eta^{\mu} = -i \rho^{a} \partial_{a}
\Pi^{\mu} ~\epsilon  ~+~\Phi^{\mu}~\epsilon ,\\
\delta \Phi^{\mu} = -i \bar{\epsilon} \rho^{a} \partial_{a}\eta^{\mu},\\
\end{array}
\begin{array}{lll}
\delta \omega = \bar{\epsilon}\xi,\\
\delta \xi^{\mu} = -i \rho^{a} \partial_{a}
\omega ~\epsilon  ~+~\Omega~\epsilon ,\\
\delta \Omega = -i \bar{\epsilon} \rho^{a} \partial_{a}\xi,
\end{array}
\eeqa
where the anti-commuting parameter $\epsilon$ is a
two-dimensional spinor \begin{eqnarray}\epsilon \equiv \left(
\begin{array}{c}
     \epsilon_{-}\\
     \epsilon_{+}
     \end{array} \right).\nonumber
\end{eqnarray}
The action (\ref{fullaction}), equations (\ref{fullequationa})
and the constraints (\ref{fullconstrants}) completely define the
system which exhibits the supersymmetry (\ref{susy}).

\section{Fermionic BRST-like symmetry}

As we shall see bellow the action (\ref{fullaction})
possesses surprisingly new global fermionic symmetry which is similar
in many ways to the BRST symmetry. This can be seen in
the light-cone coordinates. In the light-cone coordinates
the action takes the form
\beqa\label{lagran}
S = {2\over \pi} \int d^{2}\zeta ~\{ - 2\Pi^{\mu}
\partial_{+}\partial_{-} X^{\mu}
+ i\eta^{\mu}_{+}\partial_{-}\psi^{\mu}_{+}
+ i\eta^{\mu}_{-}\partial_{+}\psi^{\mu}_{-}
- {1\over 2}F^{\mu}\Phi^{\mu}~~~\nonumber\\
~~-{1\over 2}\Omega (\Pi^2 -m^2)
-  \omega (\Pi^{\mu}\Phi^{\mu} +
 i\eta^{\mu}_{+}\eta^{\mu}_{-})
-i \Pi^{\mu}~\eta^{\mu}_{+}\xi_{-}
+ i \Pi^{\mu}~\eta^{\mu}_{-}\xi_{+}~\}.
\eeqa
As one can check in addition to the SUSY transformation (\ref{susy})
this system is invariant under fermion transformation laws $\delta$
and $\bar{\delta}$:
\beqa
\begin{array}{llllll}\label{newsusy}
\delta X^{\mu} = 0,\\
\delta \Psi^{\mu}_{-} = -2\epsilon_{+}\partial_{-}X^{\mu} ,\\
\delta \Psi^{\mu}_{+}=0,\\
\delta F^{\mu} =  -2i  \epsilon_{+}   \partial_{-}\Psi^{\mu}_{+},\\
\delta \Pi^{\mu} =  i\epsilon_{+}\eta^{\mu}_{-},\\
\delta \eta^{\mu}_{-} = 0 ,\\
\delta \eta^{\mu}_{+} =  - \epsilon_{+}\Phi^{\mu}~ ,\\
\delta \Phi^{\mu} = 0,
\end{array}~~~~~
\begin{array}{lll}
\bar{\delta} X^{\mu} = 0,\\
\bar{\delta}\Psi^{\mu}_{-} = 0,\\
\bar{\delta }\Psi^{\mu}_{+}= -2\epsilon_{-}\partial_{+}X^{\mu} ,\\
\bar{\delta} F^{\mu} =  2i  \epsilon_{-}   \partial_{+}\Psi^{\mu}_{-},\\
\bar{\delta} \Pi^{\mu} =  i\epsilon_{-}\eta^{\mu}_{+},\\
\bar{\delta}\eta^{\mu}_{-} =  \epsilon_{-}\Phi^{\mu}~ ,\\
\bar{\delta}\eta^{\mu}_{+} = 0 ,\\
\bar{\delta} \Phi^{\mu} = 0,
\end{array}~~~~
\begin{array}{lll}
\delta \omega =  i\epsilon_{+}\xi_{-},\\
\delta \xi_{-} = 0,\\
\delta \xi_{+} =  -\epsilon_{+}\Omega,\\
\delta \Omega = 0,\\
\bar{\delta}\omega = i\epsilon_{-}\xi_{+},\\
\bar{\delta}\xi_{-}= \epsilon_{-}\Omega,\\
\bar{\delta}\xi_{+}=0,\\
\bar{\delta}\Omega= 0,
\end{array}
\eeqa
The  algebra  obeyed by the fermionic symmetries
is nilpotent and is very similar to BRST transformations
\be\label{brst}
\delta_{\epsilon}\delta_{\acute{\epsilon}} ~ (H)  =
\bar{\delta}_{\epsilon}\bar{\delta}_{\acute{\epsilon}}~ (H)=
0,~~~~~(\delta_{\epsilon} \bar{\delta}_{\acute{\epsilon}} -
\bar{\delta}_{\acute{\epsilon}}\delta_{\epsilon} )~ (H)=0.
\ee
where H is any of the fields $ (X,\Psi,F,\Pi,\eta,\Phi,\omega,\xi,\Omega)$.
The important fact now is that the Lagrangian is
a variation of the super-potentials $W$ and $\bar{W}$
\be
W = \Pi^{\mu}~\partial_{+}\Psi^{\mu}_{-} +
{1\over 2}\eta^{\mu}_{+}~F^{\mu},~~~~~
\bar{W} =  \Pi^{\mu}~\partial_{-}\Psi^{\mu}_{+} -
{1\over 2}\eta^{\mu}_{-}~F^{\mu},
\ee
so that
\be
\delta W = \epsilon_{+}{\cal L},~~~~~~~~~~~~~~~~~~~~~~~~~~~
\bar{\delta} ~\bar{W} =\epsilon_{-}{\cal L}.
\ee
It is also true that there exists a potential $V$
such that
\beqa
\delta ~V= -i\epsilon_{+} \bar{W},~~~~
 ~\bar{\delta} ~V =  i \epsilon_{-} W,~~~~~V= {1\over 2}\Pi^{\mu} F^{\mu},
\eeqa
thus
$$
i\delta \bar{\delta}  V = \epsilon_{+}\epsilon_{-}{\cal L},
~~~~~~~i\bar{\delta} \delta  V = \epsilon_{+}\epsilon_{-}{\cal L}.
$$
The constrains (II) can  also be represented
by the $\bar{\delta} \delta$ transformation and therefore the full
Lagrangian in (\ref{lagran}) can be represented as
\be
2\epsilon_{+}\epsilon_{-}{\cal L}_{tot} = ~i\bar{\delta} \delta~  \left(  \Pi^{\mu} F^{\mu}
+ \omega (\Pi^2 -m^2)   \right).
\ee
From (\ref{brst}) is follows that the action is invariant under
fermionic symmetries (\ref{newsusy}) and can be represented as BRST commutator
${\cal L} = \{ Q, W   \}=\{ \bar{Q},\bar{W }  \}$ where
$\delta = \epsilon_+ Q, ~\bar{\delta} = \epsilon_- \bar{Q} $,
as it takes place in topological field theories \cite{witten}.

\section{SUSY Solution}
As one can see the SUSY solution of equations (\ref{fullequationa}) is:
$$
i) \Omega =\omega  =\xi =0
$$
and the rest of the equations (I) reduce to the following form
\beqa
(I)~~~~~\partial^2 \Pi^{\mu} =0,~~~i \rho^{a} \partial_{a}\eta^{\mu}=0,~~~
\partial^2 X^{\mu} =0,~~~
i\rho^{a} \partial_{a}\Psi^{\mu} =0,~~~F^{\mu}=\Phi^{\mu}=0
\eeqa
and should be accompanied by the constraints
\beqa
(II)~~~~~\Pi^2 -m^2 =0,~~~~\bar{\eta^{\mu}}\eta^{\mu}=0,~~~~
\Pi^{\mu}~\eta^{\mu} =0.
\eeqa
In the light-cone coordinates these  equations  are easy to solve
since they take the from
\beqa
\partial_{+}\partial_{-} \Pi^{\mu} =0,~~~\partial_{\pm}\eta^{\mu}_{\mp}=0,~~~
\partial_{+}\partial_{-}  X^{\mu} =0,~~~
\partial_{\pm}\psi^{\mu}_{\mp} =0,\nonumber\\
\Pi^2 -m^2 =0,~~~~\eta^{\mu}_{+}\eta^{\mu}_{-} -
\eta^{\mu}_{-}\eta^{\mu}_{+} =0~~~~
\Pi^{\mu}~\eta^{\mu}_{\pm} =0.
\eeqa
The mass-shell supersymmetry transformation of the action
\beqa
\acute{S} = {2\over \pi} \int d^{2}\zeta ~\{ - 2\Pi^{\mu}
\partial_{+}\partial_{-} X^{\mu}
+ i\eta^{\mu}_{+}\partial_{-}\psi^{\mu}_{+}
+ i\eta^{\mu}_{-}\partial_{+}\psi^{\mu}_{-}
~\}
\eeqa
is:
\beqa\begin{array}{ll}\label{resusy}
(III)~~~\delta X^{\mu} = i\epsilon_{+}\psi^{\mu}_{-}
- i\epsilon_{-}\psi^{\mu}_{+},\\
~~~~~~~~\delta \psi^{\mu}_{-} = -2 \epsilon_{+}\partial_{-}
X^{\mu}\\
~~~~~~~~\delta \psi^{\mu}_{+} = 2 \epsilon_{-}\partial_{+}
X^{\mu}
\end{array}
\begin{array}{ll}
~~~~~~~~\delta \Pi^{\mu} = i\epsilon_{+}\eta^{\mu}_{-}
-i\epsilon_{-}\eta^{\mu}_{+},\\
~~~~~~~~\delta \eta^{\mu}_{-} = -2\epsilon_{+} \partial_{-}
\Pi^{\mu} \\
~~~~~~~~\delta \eta^{\mu}_{+} = 2\epsilon_{-} \partial_{+}
\Pi^{\mu}.
\end{array}
\eeqa
The solution of fermionic fields can be represented in
the form of mode expansion
\beqa
\eta^{\mu}_{+}= \sum c^{\mu}_{n} e^{-i n\zeta^{+}},~~~~~
\psi^{\mu}_{+}= \sum d^{\mu}_{n} e^{-i n\zeta^{+}}\nonumber\\
\eta^{\mu}_{-}= \sum \tilde{c}^{\mu}_{n} e^{-i n\zeta^{-}},~~~~~
\psi^{\mu}_{-}= \sum \tilde{d}^{\mu}_{n} e^{-i n\zeta^{-}}
\eeqa
and the basic anti-commutators should be defined as:
\beqa\begin{array}{lll}
\{ \eta^{\mu}_{\pm}(\zeta),\psi^{\nu}_{\pm}(\zeta^{'})\}= 2\pi
\eta^{\mu\nu} \delta(\zeta - \zeta^{'}),
\end{array}
\eeqa
with all others equal to zero
$\{ \eta^{\mu}_{\pm}(\zeta),\eta^{\nu}_{\pm}(\zeta^{'})\}= 0,~
\{ \psi^{\mu}_{\pm}(\zeta),\psi^{\nu}_{\pm}(\zeta^{'})\}= 0$.
Substituting the mode expansion into the anti-commutators
requires following relations between modes
\beqa\label{anticomm}
\{c^{\mu}_{n}, d^{\nu}_{k}\}= \eta^{\mu\nu}\delta_{n+k,0},~~~~
\{c^{\mu}_{n}, c^{\nu}_{k}\}=
0,~~~~~\{d^{\mu}_{n}, d^{\nu}_{k}\}= 0,
\eeqa
and similar ones for $\tilde{c}^{\mu}_{n}$ and
$\tilde{d}^{\mu}_{n}$. The commutation relations for bosonic
coordinates $X$ and $\Pi$ remain the same as in \cite{gon3}.

Our aim now is to describe the ground state sector.
Let us consider for that the fermion zero mode sector
\beqa
\{c^{\mu}_{0}, d^{\nu}_{0}\}= \eta^{\mu\nu},~~~~
\{c^{\mu}_{0}, c^{\nu}_{0}\}=
0,~~~~~\{d^{\mu}_{0}, d^{\nu}_{0}\}= 0,\\
\{\tilde{c}^{\mu}_{0}, \tilde{d}^{\mu}_{0}\}= \eta^{\mu\nu},~~~~
\{\tilde{c}^{\mu}_{0}, \tilde{c}^{\nu}_{0}\}=
0,~~~~~\{\tilde{d}^{\mu}_{0}, \tilde{d}^{\nu}_{0}\}= 0
\eeqa
together with the constraint $\eta^{\mu}_{+}\eta^{\mu}_{-} -
\eta^{\mu}_{-}\eta^{\mu}_{+} =0$, which for the ground
state takes the form $c^{\mu}_{0}\tilde{c}^{\mu}_{0} -
\tilde{c}^{\mu}_{0}c^{\mu}_{0} =0$. The nontrivial solution of the last
equations  is:
\beqa
2 c^{\mu}_{0} &= i\gamma^{\mu} \grave{\gamma}^{D+1} + \grave{\gamma}^{\mu},~~~
2 d^{\mu}_{0} = &i\gamma^{\mu} \grave{\gamma}^{D+1} - \grave{\gamma}^{\mu} ,~~~\\
2\tilde{c}^{\mu}_{0} &= \gamma^{\mu} +
i \gamma^{D+1} \grave{\gamma}^{\mu},~~~
2\tilde{d}^{\mu}_{0} = &- \gamma^{\mu} +~~
i \gamma^{D+1} \grave{\gamma}^{\mu} ,
\eeqa
where  $[\gamma^{\mu},\grave{\gamma}^{\nu}] =0$. The matrix
$\gamma^{D+1} \grave{\gamma}^{D+1}$ anticommutes  with all
these matrices and therefore the nonzero modes should be
multiplied by this matrix   to fulfill anticommutation relations
between zero modes and nonzero modes.
Thus we have the solution for
boson \cite{gon3} and fermion fields in the form
\beqa
\Pi^{\mu} &=&   m e^{\mu} +  k^{\mu}\tau + \Pi^{\mu}_{oscil},~~~
X^{\mu} =   x^{\mu} +  {1\over m}\pi^{\mu}\tau + X^{\mu}_{oscil},\\
\eta^{\mu}_{+} &=& c^{\mu}_{0} + \eta^{\mu}_{+~oscil},~~~~~~~~~~~~~~
\eta^{\mu}_{-}=  \tilde{c}^{\mu}_{0} + \eta^{\mu}_{-~oscil},
\eeqa
where $[e^{\mu}, \pi^{\nu}]=[x^{\mu}, k^{\nu}] =
i\eta^{\mu\nu}$ is a pair of conjugate variables
describing bosonic zero mode sector.
The important difference between the standard
string theory and the present one
is the appearance of conjugate variables $e$ and $\pi$,
where $e^{\mu}$ is a polarization vector
orthogonal to the momentum vector $k^{\mu}$
( $e^{\mu} k^{\mu}=0$ )~\cite{gon3} .
It is convenient to denote the ground
state wave function as $|k,e,0>$, so that
the  constraint $L_0 \Psi_{phys} = \{k \cdot \pi  +
oscillators ....\}\Psi_{phys}=0$,
which take place in this theory ( see for details \cite{gon3}), take
the form
\be\label{subsidery}
k \cdot \pi |k,e,0>= k \cdot \partial_e |k,e,0> =0.
\ee
The new constraints $\Pi \cdot\eta_{\pm}~\Psi_{phys}=0$,
which appear now in supersymmetric case, on the ground
state wave function $|k,e,0>$  will take the form
\beqa\begin{array}{ll}
(i\gamma^{\mu} \grave{\gamma}^{D+1} +
\grave{\gamma}^{\mu})k^{\mu} |k,e,0>=0\\
(i\gamma^{\mu} \grave{\gamma}^{D+1} +
\grave{\gamma}^{\mu})e^{\mu} |k,e,0>=0
\end{array}
\begin{array}{ll}
(\gamma^{\mu} + i \gamma^{D+1}
\grave{\gamma}^{\mu})k^{\mu} |k,e,0>=0\\
(\gamma^{\mu} + i \gamma^{D+1}
\grave{\gamma}^{\mu})e^{\mu}|k,e,0>=0.
\end{array}
\eeqa
Expanding the ground wave function in $e^{\mu}$ series
\be
|k,e,0> = \psi + e^{\mu} ~\psi^{\mu} + e^{\mu_1}
e^{\mu_2}~\psi^{\mu_1  \mu_2} + ...
\ee
we shall get Dirac equations on the spin tensor
$\psi^{\mu_1  \mu_2 ... \mu_J}$
\beqa
(1+ \gamma^{D+1} \grave{\gamma}^{D+1})\gamma^{\mu}
k^{\mu}~\psi^{\mu_1  \mu_2 ... \mu_J} =0\nonumber\\
(1+ \gamma^{D+1} \grave{\gamma}^{D+1})\grave{\gamma}^{\mu}
k^{\mu}~\psi^{\mu_1  \mu_2 ... \mu_J} =0
\eeqa
together with the constraint (\ref{subsidery})
\beqa
k^{\mu_1}~\psi^{\mu_1  \mu_2 ... \mu_J} =0,
\eeqa
thus describing massless particles of half-integer spin
$J+ 1/2 = 1/2, 3/2, .....$. One should study in great
details full Hilbert space of exited states
in order to learn more about complete spin content of the theory
and to prove the absence of the negative norm states.
The details will be given elsewhere.

\newpage
\section{Second Order Formulation}
In this section we shall return back to the original action
(\ref{gaga}) with its unique variable $X^{\mu}$. Using superfield
formalism one can extend
this form of the action to supersymmetric case as well. We need
just one superfield
$
\hat{X}^{\mu} =X^{\mu} + \bar{\vartheta} \Psi^{\mu}+ {1\over 2}
\bar{\vartheta} \vartheta F^{\mu}.
$
The worldsheet
supersymmecric action can be written in the form
\begin{eqnarray}\label{secondordsusy}
S = -{im \over 2\pi}
\int d^{2}\zeta d^{2} \theta \sqrt{(\bar{D} D \hat{X}^{\mu})^2}.
\end{eqnarray}
Let us demonstrate now that classically it defines a model
which is equivalent to (\ref{firstordsusy}).
Indeed the superfield equations which follow from (\ref{firstordsusy}) are:
\begin{eqnarray}
\bar{D} D \hat{X}^{\mu} - 4\hat{\Omega} \hat{\Pi}^{\mu}=0 \nonumber\\
\bar{D} D \hat{\Pi}^{\mu} =0 \nonumber\\
\hat{\Pi}^2 -m^2 =0
\end{eqnarray}
and one can see that
\begin{eqnarray}
\hat{\Pi}^{\mu} = {1\over 4\hat{\Omega}}\bar{D} D
\hat{X}^{\mu}.\nonumber
\end{eqnarray}
From the last equation it follows that
\begin{eqnarray}{1\over 4\hat{\Omega}}= {m\over \sqrt{(\bar{D} D
\hat{X}^{\mu})^2}} \nonumber
\end{eqnarray}
and field equations take the form
\begin{eqnarray}\bar{D} D \{ m {\bar{D} D \hat{X}^{\mu}\over
\sqrt{(\bar{D} D \hat{X}^{\mu})^2}}  \}=0.
\end{eqnarray}
The last equation simply follows from  (\ref{secondordsusy})
by direct variation over $\hat{X}$, thus these models are indeed classically
equivalent.

In this formulation we have
\begin{eqnarray}
(\bar{D} D
\hat{X}^{\mu})^2 = 4(F^{\mu}F^{\mu} + 2i F^{\mu} \bar{\vartheta
}\rho^{a}
\partial_{a}\Psi^{\mu} - \bar{\vartheta} \vartheta ~
F^{\mu}\partial^2 X^{\mu} - \bar{\vartheta }\rho^{a}
\partial_{a}\Psi^{\mu} \bar{\vartheta }\rho^{b}
\partial_{b}\Psi^{\mu})\nonumber
\end{eqnarray}
so that
\begin{eqnarray}
\sqrt{(\bar{D} D \hat{X}^{\mu})^2} =  2 \sqrt{F^{\mu}F^{\mu}} \{~1
- \bar{\vartheta}\vartheta ~ F^{\mu}\partial^2 X^{\mu}/2F^{2}
-(\bar{\vartheta }\rho^{a} \partial_{a}\Psi^{\mu})^2/2F^{2} +\nonumber\\
i \bar{\vartheta }\rho^{a} \partial_{a}\Psi^{\mu} F^{\mu} /F^{2}-
(1/8)(2i \bar{\vartheta }\rho^{a} \partial_{a}\Psi^{\mu} F^{\mu}
/F^{2})^2 ~\} \nonumber
\end{eqnarray}
and \begin{eqnarray}S ={m \over \pi}
\int d^{2} \zeta {1 \over \sqrt{F^{2}}} \{ F^{\mu}\partial^2
X^{\mu} + {1\over 2}
\partial_{a}\bar{\Psi}^{\mu}\rho^{a} K^{\mu\nu}
 \rho^{b} \partial_{b}\Psi^{\nu} \},~~~K^{\mu\nu}=\eta^{\mu\nu}-
 {F^{\mu} F^{\nu} \over F^2}.
\end{eqnarray}
The supersymmetry transformation is:
\begin{eqnarray}\begin{array}{ll}\label{trans}
\delta X^{\mu} = \bar{\epsilon}\Psi^{\mu},\\
\delta \Psi^{\mu} = -i \rho^{a} \partial_{a}
X^{\mu} ~\epsilon  ~+~F^{\mu}~\epsilon ,\\
\delta F^{\mu} = -i \bar{\epsilon} \rho^{a}
\partial_{a}\Psi^{\mu}.
\end{array}
\end{eqnarray}
We can write the action  in terms of components in the
following form
\begin{eqnarray}\label{19}
S=\frac{2m}{\pi}\int d^{2}\zeta
\frac{1}{\sqrt{F^{2}}}\left\{-2F^{\mu}\partial_{+}\partial_{-}X^{\mu}
+i\left[\partial_{-}\Psi_{+}^{\mu}\partial_{+}
\Psi_{-}^{\nu} -  \partial_{+}
\Psi_{-}^{\mu}\partial_{-}\Psi_{+}^{\nu}\right]K^{\mu\nu}\right\}
\end{eqnarray}
with the following equations of motion for
$F^{\mu},\Psi^{\mu}_{\pm}$ and $X^{\mu}$:
\begin{eqnarray}\label{20}
&&K^{\mu\nu}\partial_{+}\partial_{-}X^{\nu}-\frac{i}{2}\left[\partial_{+}
\Psi_{-}^{\nu}\partial_{-}\Psi_{+}^{\lambda}-\partial_{-}
\Psi_{+}^{\nu}\partial_{+}\Psi_{-}^{\lambda}\right]
\frac{F_{\{\mu}K_{\nu\lambda\}}}{F^{2}}=0,\\
&&\partial_{\pm}\left[\frac{1}{\sqrt{F^{2}}}K_{\mu\nu}\partial_{\mp}
\Psi^{\nu}_{\pm}\right]=0,
\label{21}\\
&&\partial^{2}\left(\frac{F^{\mu}}{\sqrt{F^{2}}}\right)=0\label{22},
\end{eqnarray}
where $\{\mu\nu\lambda\}=\mu\nu\lambda+ \rm{cycl. perm.}$. Solving
Eq. (\ref{21}) and using (\ref{20}) we shall get the following
solution and constraints for fermions
\begin{eqnarray}\label{23}
  &&\Psi^{\mu}_{+}(\zeta^{+},\zeta^{-})=\psi^{\mu}_{+}(\zeta^{+})+
  \int^{\zeta^{-}}_{0}
  d\tilde{\zeta}^{-}\left(\frac{1}{m}\eta^{\mu}_{-}(\tilde{\zeta^{-}})
\Lambda(\zeta^{+}\tilde{\zeta^{-}})-\frac{1}{2}\Pi^{\mu}(\zeta^{+}\tilde{\zeta^{+}})
\xi_{-}(\zeta^{+}\tilde{\zeta^{-}})\right)\qquad\\
&&\Psi^{\mu}_{-}(\zeta^{+},\zeta^{-})=\psi^{\mu}_{-}(\zeta^{-})+
\int^{\zeta^{+}}_{0}
d\tilde{\zeta}^{+}\left(\frac{1}{m}\eta^{\mu}_{+}(\tilde{\zeta^{+}})
\Lambda(\tilde{\zeta^{+}}\zeta^{-})-\frac{1}{2}\Pi^{\mu}(\tilde{\zeta^{+}}\
zeta^{+})\xi_{+}(\tilde{\zeta^{+}}\zeta^{-})\right)\qquad\\
&&\Pi^{\mu}\eta^{\mu}_{\pm}=0,\qquad
\eta^{\mu}_{+}\eta^{\mu}_{-}-\eta^{\mu}_{-}\eta^{\mu}_{+}=0
\label{24}
\end{eqnarray}
where we have introduced instead of vector  $F^{\mu}$ the normalized
variable $\Pi^{\mu}=m\frac{F^{\mu}}{\sqrt{F^2}}$ and scalar
variable $\Lambda=\sqrt{F^{2}}$ corresponding to the length of
$F^{\mu}$. The remaining equations in the bosonic sector are the same
as in the ``first'' order formalism  described in the
previous sections
\begin{eqnarray}\label{25}
  &&\partial_{+}\partial_{-}\Pi^{\mu}=0,\qquad \Pi^{2} - m^2=0,\\
&&\partial_{+}\partial_{-}X^{\mu}+\frac{1}{2}\Omega\Pi^{\mu}
+\frac{i}{2}\left(\eta^{\mu}_{+}\xi_{-}-\eta_{-}\xi_{+}\right)=0.
\end{eqnarray}
From the action (\ref{19}) we can derive the
canonical momenta for  fermionic fields
\begin{eqnarray}\label{momenta}
P_{\Psi_{-}}^{\mu}=\frac{\overrightarrow{\delta} L}{\delta
\partial_{0}\Psi_{-}^{\mu}}=-\frac{2m}{\pi\sqrt{F^{2}}}K_{\mu\nu}\partial_{-
}\Psi_{+}\\
P_{\Psi_{+}}^{\mu}=\frac{\overleftarrow{\delta} L}{\delta
\partial_{0}\Psi_{+}^{\mu}}=-\frac{2m}{\pi\sqrt{F^{2}}}K_{\mu\nu}\partial_{+
}\Psi_{-}
\end{eqnarray}
Using solutions (\ref{23}) and (\ref{24}) we can  see that these
momenta are equal to $\eta^{\mu}_{\pm}$,  and anti-commutators between
$\eta$ and $\psi$ coincide with anti-commutation relations
(\ref{anticomm}) of the previous section.

The advantage of the last formalism is that
it has less fields and therefore some hidden
symmetries of the model  are much easier to detect.
In particular, it is much easier to observe a surprising
property of the action (\ref{19}), that it is BRST
exact and might be related to the topological nature of this
supersymmetric extension. More
precisely we can define the following set of nilpotent Grassmann
odd  symmetries related with some part of supersymmetry
transformations
\begin{eqnarray}\label{36}
&&{\bf S}\Psi^{\mu}_{+}=\partial_{+}X^{\mu}, \qquad
{\bf S}X^{\mu}=0,\nonumber\\
&&{\bf S}F^{\mu}=-i\partial_{+}\Psi^{\mu}_{-}, \qquad
{\bf S}\Psi^{\mu}_{-}=0,\label{37}\nonumber\\
&&{\mathbf S}^{2}=0\label{38}
\end{eqnarray}
or the similar in left moving sector
\begin{eqnarray}\label{39}
&&{\bf \bar{S}}\Psi^{\mu}_{-}=\partial_{-}X^{\mu}, \qquad
{\bf \bar{S}}X^{\mu}=0,\nonumber\\
&&{\bf \bar{S}}F^{\mu}=i \partial_{-}\Psi^{\mu}_{+}, \qquad
{\bf \bar{S}}\Psi^{\mu}_{+}=0,\label{40}\nonumber\\
&&{\bf \bar{S}}^{2}=0\label{41}
\end{eqnarray}
The action (\ref{19}) can be expressed in the BRST exact form
\begin{eqnarray}\label{42}
S=\frac{-4m}{\pi}\int d^{2}\zeta {\bf S} W= \frac{-4m}{\pi} \int
d^{2}\zeta{\bf \bar{S}} \bar{W}
,\end{eqnarray}
where we have introduced so called gauge fermions
\begin{eqnarray}\label{43}
W= \partial_{-}\Psi^{\mu}_{+}\frac{F^{\mu}}{\sqrt{F^{2}}},~~~~~~~~~~~
\bar{W}= \partial_{+}\Psi^{\mu}_{-}\frac{F^{\mu}}{\sqrt{F^{2}}}.
\end{eqnarray}
Then we can observe that these two nilpotent symmetries anticomute
\begin{eqnarray}\label{44}
\left\{{\mathbf S},{\bf \bar{S}}\right\}=0
\end{eqnarray}
and we can express our two gauge fermions in term of one gauge
boson and generators of the second symmetry
\begin{eqnarray}\label{45}
W= {\bf \bar{S}}\sqrt{F^{2}},~~~~~~~~~~~~~~~~~~~~~~~~~~
\bar{W}=-{\mathbf S}\sqrt{F^{2}},\\
S=\frac{-4m}{\pi}\int d^{2}\zeta~
{\bf S}{\bf\bar{S}}~\sqrt{F^{2}}.~~~~~~~~~~~~~~~~~
\end{eqnarray}
So we can deduce that the action (\ref{19}) is BRST and anti-BRST
exact and somehow is equivalent to the sum of gauge fixing term and
Faddeev-Popov determinant for some  gauge transformation of
our fields (may be in this case nonlocal).

One can formally express the partition function
in terms of singular determinant, if for a moment
ignor its zero modes. Indeed, integrating in Z
\begin{eqnarray}\label{46}
  Z=\int \exp\left[iS\right] DF^{\mu}D\Psi^{\mu}_{-}D\Psi^{\mu}_{+}D\Psi^{\mu}_{-}
\end{eqnarray}
over $\Psi^{\mu}_{\pm}$ and $X^{\mu}$ we shall get the following
expression
\begin{eqnarray}\label{47}
  Z=\int \det\left[\partial_{-}\frac{\delta \Pi^{\mu}}{\delta
  F^{\mu}}\partial_{+}\right]
  \delta\left(\partial_{+}\partial_{-}\Pi^{\mu}\right)DF^{\mu},
\end{eqnarray}
where $\Pi^{\mu}=\frac{F^{\mu}}{\sqrt{F^{2}}}$ and $\frac{\delta
\Pi^{\mu}}{\delta   F^{\nu}}=\frac{K^{\mu\nu}}{\sqrt{F^{2}}}$.
It is clear that if one changes the integration variable
$F^{\mu}$ to a unit length function $\Pi^{\mu}$ then two determinants
will cancel each other and the partition
function Z will be equal to one. But we have to notice that this
transformation is degenerate because the corresponding Jacobian
$\frac{\delta \Pi^{\mu}}{\delta F^{\mu}}$ is proportional to
the projector $K^{\mu\nu}$ and therefore one should properly account
its zero modes.

As usually in the case of BRST exact actions \cite{witten},
we can construct  observables. For any form
$A=A\left(F\right)_{\mu_{1}....\mu_{n}}dF^{\mu_{1}}\wedge....\wedge
dF^{\mu_{n}}$ we can define  the corresponding
operator ${\cal O}_{A}$ replacing $dF^{\mu_{i}}$
by the BRST  exact operator $\partial_{-}\Psi^{\mu_{i}}_{+}$
\begin{eqnarray}\label{48}
{\cal O}_{A}=A\left(F \right)_{\mu_{1}....\mu_{n}}\partial_{-}
\Psi^{\mu_{1}}_{+}~...~\partial_{-}\Psi^{\mu_{n}}_{+}.
\end{eqnarray}
From (\ref{36}) it follows that
\begin{eqnarray}\label{49}
&&i\bar{S}{\cal O}_{A}=
{\partial A_{\mu_{1}....\mu_{n}}  \over \partial F^{\mu_{0}} }
~~\partial_{-}\Psi^{\mu_{0}}_{+}~\partial_{-}
\Psi^{\mu_{1}}_{+}~...~\partial_{-}\Psi^{\mu_{n}}_{+}\\
&&i\bar{S}{\cal O}_{A}={\cal O}_{dA}
\end{eqnarray}
where $dA$ is the exterior derivative of A. Thus closed forms
$dA =0$ induce BRST invariant operators ${\cal O}_A$ and de Rham cohomology classes
of forms transform to the BRST cohomology classes, and we
have at our disposal a set of nontrivial invariant observables
\cite{witten}.

\vfill
\end{document}